\documentclass[review]{elsarticle}

\usepackage{multirow}
\usepackage{subfigure}
\usepackage{booktabs}
\usepackage{amsmath,xparse}

\journal{Journal of \LaTeX\ Templates}

\begin{document}

\begin{frontmatter}

\title{Dual-Domain Fusion Convolutional Neural Network for Contrast Enhancement Forensics \tnoteref{mytitlenote}}
\tnotetext[mytitlenote]{Fully documented templates are available in the elsarticle package on \href{http://www.ctan.org/tex-archive/macros/latex/contrib/elsarticle}{CTAN}.}


\author[mymainaddress,mysecondaddress]{Pengpeng Yang}
\author[mymainaddress,mysecondaddress]{Rongrong Ni\corref{mycorrespondingauthor}}
\cortext[mycorrespondingauthor]{Corresponding author}
\ead{rrni@bjtu.edu.cn}

\author[mymainaddress,mysecondaddress]{Yao Zhao}
\author[mythirdaddress]{Gang Cao}
\author[mymainaddress,mysecondaddress]{Wei Zhao}
\address[mymainaddress]{Institute of Information Science, Beijing Jiaotong University, Beijing, {\rm 100044}, CHINA}
\address[mysecondaddress]{Beijing Key Laboratory of Advanced Information Science and Network Technology, Beijing {\rm 100044}, CHINA}
\address[mythirdaddress]{Communication University of China, Beijing {\rm 100024}, CHINA}




\begin{abstract}
Contrast enhancement (CE) forensics techniques have always been of great interest for image forensics community, as they can be an effective tool for recovering image history and identifying tampered images. Although several CE forensic algorithms have been proposed, their accuracy and robustness against some kinds of processing are still unsatisfactory. In order to attenuate such deficiency, in this paper we propose a new framework based on dual-domain fusion convolutional neural network to fuse the features of pixel and histogram domains for CE forensics. Specifically, we first present a pixel-domain convolutional neural network (P-CNN) to automatically capture the patterns of contrast-enhanced images in the pixel domain. Then, we present a histogram-domain convolutional neural network (H-CNN) to extract the features in the histogram domain. The feature representations of pixel and histogram domains are fused and fed into two fully connected layers for the classification of contrast-enhanced images. Experimental results show that the proposed method achieve better performance and is robust against pre-JPEG compression and anti-forensics attacks. In addition, a strategy for performance improvement of CNN-based forensics is explored, which could provide guidance for the design of CNN-based forensics tools. 
\end{abstract}


\end{frontmatter}

\section{Introduction}
Being a simple yet efficient image processing operation, CE is typically used by malicious image attackers to eliminate inconsistent brightness when generating visually imperceptible tampered images. CE detection algorithms play an important role in decision analysis for authenticity and integrity of digital images. Although some schemes have been proposed to detect contrast-enhanced images, the performance of such techniques is limited in the cases of pre-JPEG compression and anti-forensic attacks. Therefore, it is critical to develop robust and effective CE forensics algorithms. 

Thanks to the efforts of researches in the past decade, a number of schemes \cite{stamm2008blind, stamm2010forensic, stammConf, cao2014contrast, li2016identification, lin2013exposing, lin2014two, wen2018contrast, de2015second} has been proposed to discriminate  contrast-enhanced images in uncompressed format. Stamm \emph{et al.} \cite{stamm2008blind, stamm2010forensic, stammConf} found that contrast enhancement would introduce peaks and gaps into the image's gray level histogram, which led to specific high values in high-frequency components. Lin \emph{et al.} \cite{lin2013exposing, lin2014two} revealed that contrast enhancement would disturb the inter-channel correlation left by color image interpolation and they measured such correlation to distinguish the enhanced images from the original images. Furthermore, in order to recover the image processing history, many algorithms for estimating parameters for contrast-enhanced images have been developed \cite{farid2001blind, popescu2004statistical, cao2010forensic, wang2018parameter}.

Despite the good performance obtained by the abovementioned algorithms, their robustness can be unsatisfactory in some cases, such as the CE of JPEG images (pre-JPEG compression) and the occurrence of anti-forensic attacks \cite{barni2012universal, cao2010anti,kwok2011alternative, comesana2013optimal, cao2014attacking, ravi2015ace}. The reason lies in that the fingerprint left by CE operation would be altered. Based on such a phenomenon, some researchers proposed more robust CE forensic algorithms, which can be divided into two major branches: overcoming pre-JPEG compression \cite{cao2014contrast} and defending against anti-forensic attacks \cite{de2015second}. Unfortunately, neither one of these methods is capable of addressing both pre-JPEG compression and anti-forensic attacks. To date there are no satisfactory solutions for these problems. 

With the rapid development of deep learning technique, and especially convolutional neural networks (CNNs), some researchers have recently attempted to use them for digital image forensics. A number of preliminary works exploring CNNs in a single-domain (such as the pixel domain\cite{barni2018cnn}, the histogram domain\cite{zhang2018global}, and the gray-level co-occurrence matrix (GLCM) \cite{sun2018novel, shan2019robust}) has been proposed for CE forensics. According to the report \cite{sun2018novel}, deep learning-based CE forensic schemes have achieved better performance than traditional ones. The schemes mentioned above try to deal with CE forensics task by feeding single-domain information to CNNs. However, each domain has its own advantages and disadvantages. For example, according to our experiments, the CNN working in the pixel domain is robust to post-processing but hard to get satisfactory performance. In addition, it is well known that histogram domain is effective for CE forensics task but fails to resist to CE attacks. Such situations give us strong incentive to explore fusion algorithm across multiple domains based on deep learning technique against pre-JPEG compression and anti-forensic attacks. 

In this paper, we propose a novel framework based on dual-domain fusion convolutional neural network for CE forensics. Specifically, pixel-domain CNN (P-CNN) is designed for the pattern extraction of contrast-enhanced image in pixel domain. For P-CNN, high-pass filter is used to reduce the affect of image contents and keep the data distribution balance cooperating with batch normalization \cite{ioffe2015batch}. In addition, the histogram-domain CNN (H-CNN) is constructed by feeding an histogram with 256 dimensions into convolutional neural network. The features obtained from P-CNN and H-CNN are fused together and fed into a classifier with two fully connected layers. Experimental results show that our proposed method outperforms  state-of-the-art schemes in the case of uncompressed images and obtains comparable performance in the cases of pre-JPEG compression, anti-forensics attack, and CE level variation.

The main contributions of this paper are:

1) we present a dual-domain fusion framework for CE forensics;

2) we propose and evaluate two kinds of simple yet effective convolutional neural networks based on pixel and histogram domains;

3) we explore the design principle of CNN for CE forensics, specifically, adding the preprocessing, improving complexity of architecture, and selecting training strategy that includes fine-tune technique and data augmentation.

The rest of this paper is organized as follows. Section 2 describes related works in the field of CE forensics. In Section 3, we formulate the problem and in Section 4 we present the proposed dual-domain fusion CNN framework. In Section 5, experimental results are reported. Conclusion is given in Section 6.

\section{Related Works}
CE forensics, as a popular topic in image forensics community, has been study for a long time. Early research works attempt to extract features from the histogram domain. Stamm \emph{et al.} \cite{stamm2008blind, stamm2010forensic, stammConf} observed that the histogram of contrast-enhanced images presents peak/gaps artifacts, in contrast, that of un-enhanced image does not occur the peak/gaps, as shown in Fig 1. Based on such observation, they proposed the histogram-based scheme that the high frequency energy metric is calculated and decided by threshold strategy. However, the above method failed to detect CE image in previously middle/lower quality JPEG compressed images in which the peak/gaps artifacts also exits \cite{cao2014contrast}. Cao \emph{et al.} \cite{cao2014contrast} studied this issue and found that there exists notable difference between the peak/gap artifacts from contrast enhancement and those from JPEG compression, which is that the gap bins with zero height always appear in contrast-enhanced images. But the above phenomenon does not occur in the case of anti-forensics attack. As can be seen in Fig 1, the histogram of enhanced image with anti-forensics attack conforms to a smooth envelope, which is similar with the un-enhanced image.

\begin{figure}[!h]
	\centering
	\includegraphics[width=13cm,height=10cm]{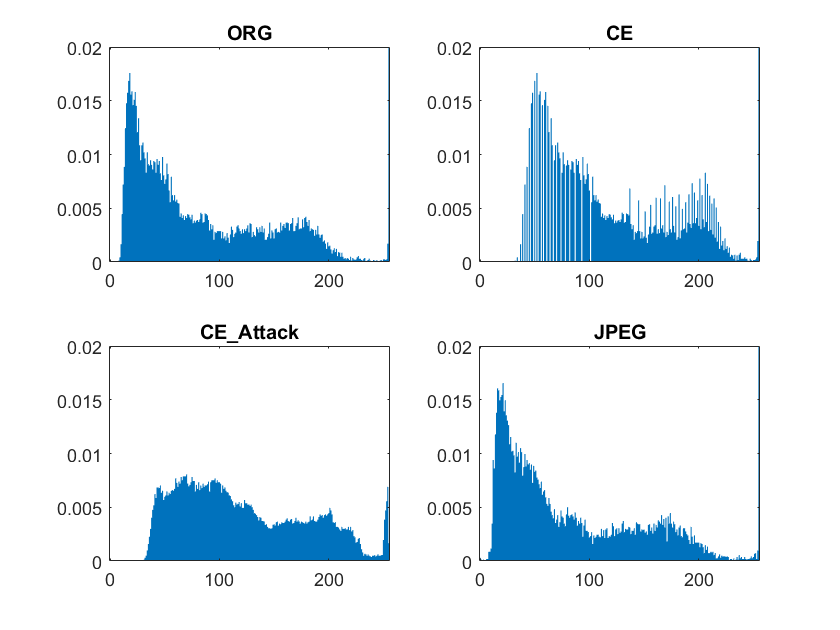}
	\caption{Histogram of uncompressed image, contrast enhanced image with $\gamma=0.6$, contrast enhanced image in the case of anti-forensic attack, JPEG image that quality factor is equal to 70, respectively.}
\end{figure}

Instead of exploring the features in histogram domain, De Rosa \emph{et al.} \cite{de2015second} studied the possibility of using second order statistics to detect contrast-enhanced images even in the case of anti-forensics attack. Specifically, the co-occurrence matrix of a gray-level image was explored. According to the report\cite{de2015second}, several empty rows and columns appears in the GLCM of contrast-enhanced images, as shown in Fig 2, even after the application of anti-forensics attack\cite{barni2012universal}. Based on this observation, the authors tried to extract such feature from the standard deviation of each column of the GLCM. However, its performance still not satisfactory, especially for the other powerful anti-forensics attack\cite{cao2010forensic}.

These algorithms described are based on handcrafted low-level features which is not easy to deal with the above problems simultaneously. With the development of data-driven technique, some researchers have started to study the deep feature represents for CE forensics via data-driven approach recently and existing methods \cite{sun2018novel, barni2018cnn, zhang2018global, shan2019robust} focus on exploring in single-domain. Barni \emph{et al.} \cite{barni2018cnn} present a CNN containing a total of 9 convolutional layers in pixel domain which is similar with the typical CNNs used in the field of computer vision. Cong \emph{et al.} \cite{zhang2018global} explore the information in histogram domain and apply the histogram with 256 dimensions into VGG-based multi-path network. Sun \emph{et al.} \cite{sun2018novel} propose to calculate the gray-level co-occurrence matrix (GLCM) and feed it to a CNN with 3 convolutional layers. Although these approaches based on deep features in single-domain have obtained performance gain for CE forensics, they ignore multi-domain information which could be useful in the case that some features in single-domain are destroyed. 

To overcome these limitation of exiting works, we propose a new deep learning-based framework to extract and fuse feature representation in pixel and histogram domains for CE forensics.

\begin{figure}[!h]
	\centering
	\includegraphics[width=13cm,height=10cm]{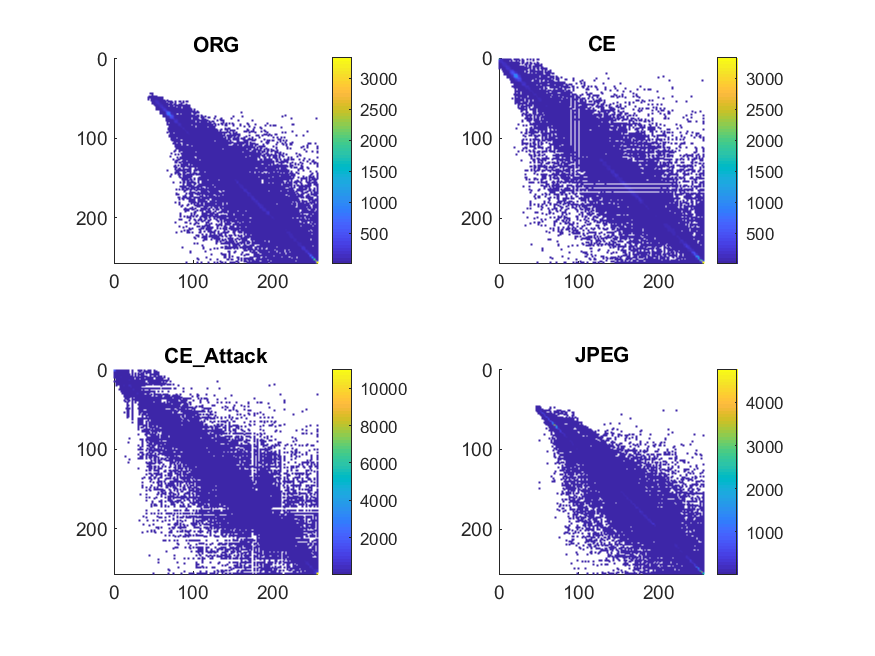}
	\caption{GLCM of uncompressed image, contrast enhanced image with $\gamma=0.6$, contrast enhanced image in the case of anti-forensic attack, JPEG image that quality factor is equal to 70, respectively.}
\end{figure}


\section{Problem Formulation}
As a common way of contrast enhancement, gamma correction can be found in many image-editing tools. In addition, according to the report\cite{barni2018cnn}, enhanced-images with gamma correction is harder to be detected than the enhance-images via the other way. Therefore, in this paper, we mainly focus on the detection of gamma correlation, which is typically defined as,
\begin{equation}\label{key}
Y=[255(X/255)^\gamma ]\approx 255(T^\gamma)  
\end{equation}
where $X$ denotes an input and $Y$ represents the re-mapped value, $T=(X/255) \epsilon [0,1]$. 
The problem addressed in this paper is how to classify the given image as contrast enhanced or non-enhanced image. Particularly, the robustness of proposed method against pre-JPEG compression and anti-forensics attacks is evaluated.

\section{Proposed Method}
In this section, we first make an overview of the proposed framework dual-domain fusion convolutional neural network, and then introduce the major components in detail.
\subsection{Framework Overview}
The proposed dual-domain fusion convolutional neural network is shown in Fig 3, which extracts the features from pixel and histogram domains by P-CNN and H-CNN, respectively, and then fuses them before feeding into the classifier with two fully-connected layers. Our end-to-end system would predict whether the image is a contrast enhanced or non-enhanced image.

\begin{figure*}[t]
	\centering
	\includegraphics[width=15cm,height=9cm]{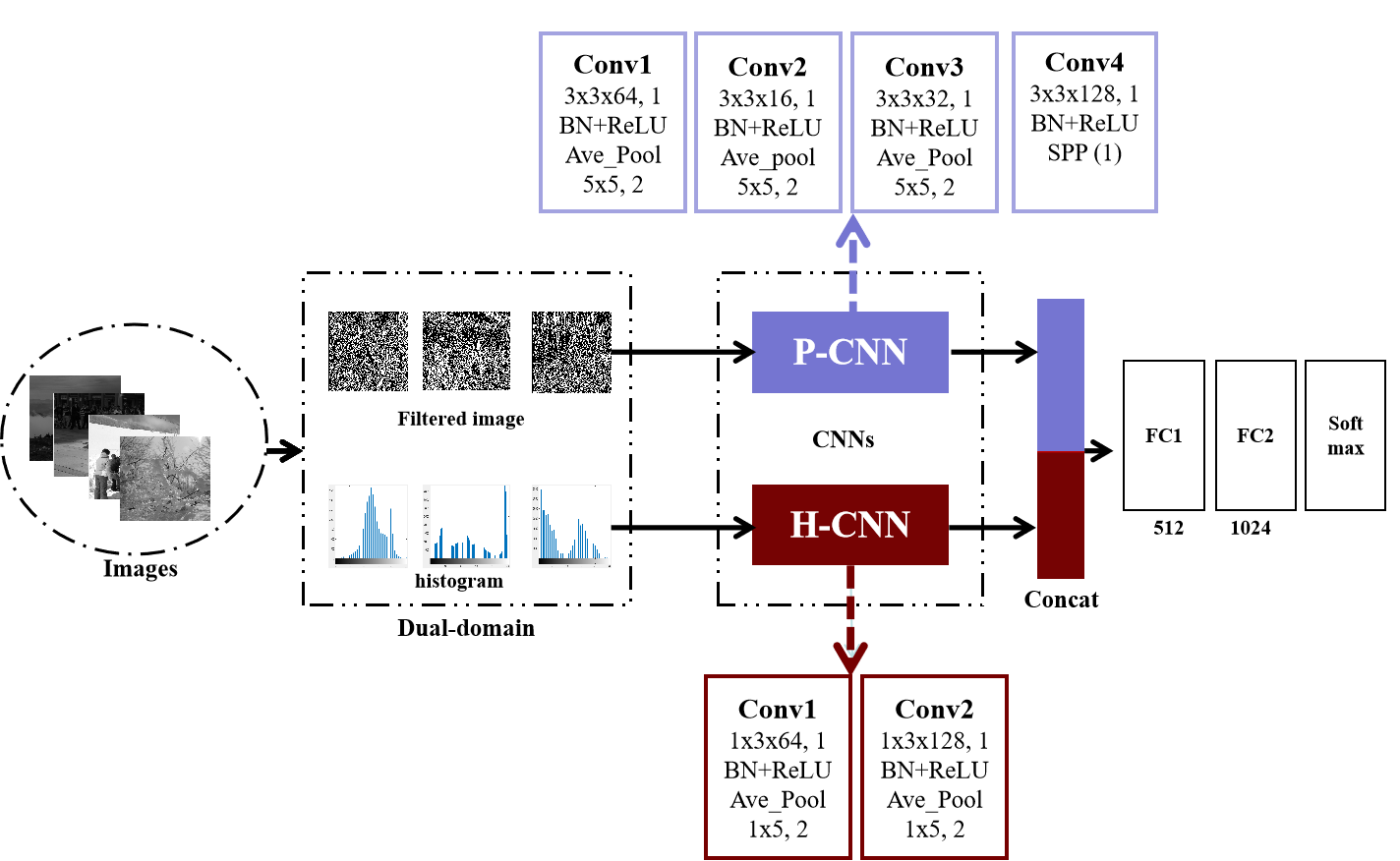}
	\caption{The proposed dual-domain fusion convolutional neural network.}
\end{figure*}
\subsection{Pixel-Domain Convolutional Neural Network}
Convolutional neural networks (CNNs) in pixel domain have been applied in image forensics and developed for specific forensic tasks recently.  The common modification \cite{yang2016recapture, yang2017source} for the CNNs in forensics community is to add preprocessing layer that could weaken the effect of image content and improve the signal noise ratio. Inspired by this observation, we experimental study on preprocessing and find effective way for CE forensics (Section 5.3.1). Due to the limitation of hardware, we design a simple 4 layers CNN to keep the balance between performance and computational complexity. The architecture of proposed pixel-domain convolutional neural network is shown in Fig 4.

\begin{figure}[t]
	\includegraphics[width=15cm,height=4cm]{{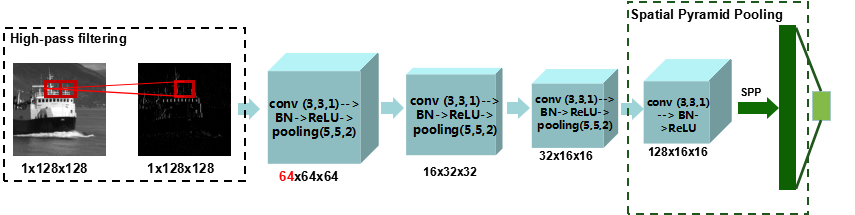}}
	\caption{The architecture of proposed pixel-domain convolutional neural networks.}
\end{figure}

Firstly, the high-pass filter is added into the front-end of architecture to eliminate the interfere of image content. Another advantage of using high-pass filter could be that it accelerates training by cooperating with batch normalization. Because that the histogram of high-pass filtered images approximately follows the generalized Gaussian distribution, which is similar to batch normalization \cite{ioffe2015batch}. In particular, we experimentally find that the filter of the first-order difference along horizontal direction has better performance. 

\begin{equation}\label{key}
I_{1} = H*I
\end{equation}
where $H=[1,-1]$, $I$ is the input image, $I_{1}$ is the output of the first layer, '*' represents the convolution operator.

Next, high-pass filtering layer is followed by four traditional convolutional layers. For each layer, there are four types of operations: convolution, batch normalization, ReLU and average pooling. The feature maps for each layer are 64, 16, 32, 128, respectively. The kernel size for convolutional and pooling operation is 3x3 with 1 stride, 5x5 with 2 strides. It should be pointed out that: 1) we experimentally find that the numbers of feature map for first convolutional layer is important for CE detection and it has better performance when the feature maps is 64. In other words, low-level feature would be more helpful; 2) instead of average pooling, the spatial pyramid pooling (SPP) layer \cite{he2015spatial} is used in last convolutional layer to fuse multi-scale features. The convolutional layer is calculated as

\begin{equation}\label{key}
I_{i}=
\left\{
\begin{array}{lr}
P(R(F(W_{i}*I_{i-1}+B_{i}))), i\epsilon (2,3,4)\\ 
S(R(F(W_{i}*I_{i-1}+B_{i}))), i=5
\end{array}
\right.
\end{equation}
where $F, R, P ,S$ represent the batch normalization, ReLU, average pooling, and spatial pyramid pooling, respectively. For spatial pyramid pooling, three scales are chosen and lead to 2688 dimensional output.

In the end, the fully connected layer and softmax is followed by a multinomial logistic loss. The loss function is defined as,
\begin{equation}\label{key}
Loss=-log(\frac{e^{W^{j}I_{5}+B^{j}}}{\sum_{j=1}^{n}e^{W^{j}I_{5}+B^{j}}})
\end{equation}
where $n$ is the number of classes and $j$ denotes the true label. In our experimental setup, Mini-batch Stochastic Gradient Descent is applied and the batch size is set as 120. The learning rate is initialized as 0.001, and scheduled to decrease 10\% for every 10000 iterations. The max iterations is 100000. The momentum and weight\_decay are fixed to 0.9 and 0.0005, respectively.

\subsection{Histogram-Domain Convolutional Neural Network}
\begin{figure}[h]
	\centering
	\includegraphics[width=14cm,height=4.3cm]{{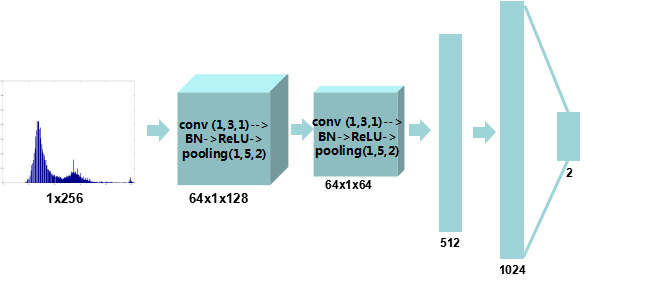}}
	\caption{The architecture of proposed histogram-domain convolutional neural networks.}
\end{figure}
 
 As well known, gamma correction would lead to the non-linear changes in pixel domain and introduce the peak/gap bins into histogram domain \cite{stamm2008blind, stamm2010forensic, stammConf, cao2014contrast}. A number of handcrafted features have been designed based on such phenomenons. Instead of designing features, the histogram-domain convolutional neural networks (H-CNN) is constructed to achieve end-to-end self-learning detection. The H-CNN is proposed to self-learn better feature directly from histogram domain. In addition, as an input with low and fixed dimension, the histogram is suitable for convolutional neural networks. The architecture of H-CNN is shown in Fig 5. Its input is the histogram of the image, namely a vector with 1x256 dimensions. Then, such an input layer is followed by two convolutional and two fully connected layers. The feature maps are 64, 64, 512, 1024, respectively. Lastly, the softmax layer followed by a multinomial logistic loss is added to classify original and enhanced images. The parameters of convolutional layers and hyper-parameters are the same as the P-CNN.

\subsection{Dual-domain Fusion Convolutional Neural Network}
According to the description in Section 1,2, the performance of CE system designed in single-domain is still not satisfactory. Fortunately, fusion strategies \cite{mangai2010survey} provide a good solution to obtain higher performance and have been adopted in the community of digital image forensics \cite{yang2017source, fontani2013framework}. In this work, we assume that the features extracted from P-CNN and H-CNN are complementary for CE forensics, thus we propose a simple yet effective feature fusion framework for deep learning-based CE forensics to integrate multiple domains and construct the dual-domains fusion CNN (DM-CNN), as shown in Fig 3. Firstly, high-pass filtered images and the histogram are extracted from input images. Then the filtered images are fed into P-CNN with four 2D-convolutional layers and the histogram is fed into H-CNN with two 1D-convolutional layers. Note that for the purpose of fusion, P-CNN and H-CNN are slightly modified. The P-CNN of DM-CNN is composed of the convolutional layers extracted from the P-CNN. Besides, in order to ensure that the outputs of the P-CNN and H-CNN have the same dimension, one scale of spatial pyramid pooling in P-CNN is chosen and the number of feature map in the second convolutional layer of H-CNN is set to 128. The features output from of P-CNN and H-CNN are concatenated together and then fed into classification unit, which consists of two fully connected layers and one softmax layer followed by multinomial logistic loss. It is worth noting that due to the limitation of our hardware configuration, only dual-domains are fused in our system and it would be useful to ensemble features from the other domains.    

\section{Experimental Results}
In order to verify the validity of proposed methods, we compared them with four other methods. De Rosa \cite{de2015second}, Cao \cite{cao2014contrast} and Sun \cite{sun2018novel} are proposed for CE forensic. The former two algorithms belong to traditional scheme and the last one is based on deep learning technique. Li \cite{li2016identification} is proposed to identify various image operations using high-dimensional residual-based features. Four groups of experiments are conducted: ORG vs P\--CE, JPEG\--ORG vs JPEG\--CE, ORG vs Anti\--CE, and JPEG\--ORG vs JPEG\--CE\--Anti\--CE,  where ORG is original images in uncompressed format, JPEG\--ORG represents original images in JPEG format, P\--CE and JPEG\--CE denote enhanced versions of ORG and JPEG\--ORG, respectively, and Anti\--CE and JPEG\--CE\--Anti\--CE represent enhanced images with anti-forensics attack for P\--CE, JPEG\--CE, respectively. The BOSSBase \cite{Bossbasedata} with 10000 images is chosen to construct the dataset. Firstly, the images are centrally cropped into 128x128 pixel patches as ORG. Then, JPEG compression with $Q=70, 50$ is carried out for ORG to build JPEG\--ORG. Next, gamma correction with $\gamma=\{0.6, 0.8, 1.2, 1.4\}$ is implemented on ORG, JPEG\--ORG to constitute P\--CE and JPEG\--CE. In the end, Anti\--CE is produced by anti-forensics attacks \cite{cao2010forensic, barni2012universal} on P\--CE and JPEG\--CE. The reasons for our choice of pixel patch size are that 1) the detection for the images with lower resolution is much harder than higher resolution image; 2) 128x128 is a suitable size for tamper locating based on CE forensics; 3) our hardware configuration is limited. For each experiment, the training data, validation and testing data is 8000, 2000, 10000, respectively. The experiments about the proposed schemes are conducted on one GPU (NVIDIA TITAN X) with an open source framework of deep learning: Caffe \cite{caffewebsit}.

\subsection{Contrast Enhancement Detection: ORG vs P\-CE }
The result for contrast-enhanced images in uncompressed format, is as shown in Table 1. P-CNN is pixel-domain convolutional neural networks and H-CNN is histogram-domain convolutional neural networks. DM-CNN denotes the dual-domain fusion CNN. As seen from the Table 1, for Cao's method, the detection accuracy for $\gamma=\{0.6, 0.8\}$ is much higher than one for $\gamma=\{1.2, 1.4\}$. The reason is that gap feature is unstable among CE parameters, which is consistent with our analysis in Section III. In addition, H-CNN has better performance than the above four schemes.  Such results demonstrated that the histogram domain feature should be effective for CE detection. Besides, proposed fusion framework, DM-CNN, obtains best average detection accuracy. It should be mentioned that although the deep learning-based method proposed by Sun obtained slightly lower detection accuracy than DM-CNN, it has a much higher computational cost during the feature extraction of the GLCM in preprocessing.

\begin{table}[!h]
	\footnotesize
	\caption{CE detection accuracy for contrast-enhanced images in the case that ORG vs P\--CE. AVE is the average accuracy. Best results are marked in bold.}
	\label{tab1}
	\tabcolsep 13pt 
	\begin{tabular*}{\textwidth}{c|cccc|c}
		\toprule
		Method & $\gamma=0.6$    &  $\gamma=0.8$  & $\gamma=1.2$ & $\gamma=1.4$ & AVE \\\hline
		De Rosa\cite{de2015second} & 94.02\% & 84.85\% & 78.37\% & 74.12\% & 82.84\%\\
		Cao\cite{cao2014contrast} & 93.89\% & 93.90\% & 80.26\% & 81.40\% & 87.36\%\\
		Li\cite{li2016identification} &  93.63\% & 89.48\% & 90.76\% & 93.44\%  & 91.83\%\\
		Sun\cite{sun2018novel} & 99.35\%  & 99.21\% & 98.45\% & 98.80\% & 98.95\%\\
		P-CNN & 94.70\%  & 89.00\% & 78.00\% & 86.00\% & 86.93\%\\
		H-CNN & 99.48\%  & 99.45\% & \textbf{99.40\%} & 99.07\% & 99.35\%\\
		DM-CNN & \textbf{99.80\%} & \textbf{99.72\%} & 99.36\% & \textbf{99.41\%} & \textbf{99.57\%}\\
		\bottomrule
	\end{tabular*}
\end{table}

\subsection{Robustness Against Pre-JPEG Compressed and Anti-Forensic Attacked Contrast-Enhanced Images}
The performance of different methods for pre-JPEG compressed images with $Q=\{50, 70\}$ and anti-forensics attacked images are shown in Table 2, 3, 4. It can be seen from Table 2 that P-CNN, H-CNN, DM-CNN have much higher detection accuracy than De Rosa's and Cao's methods and comparable performance with the algorithms proposed by Li and Sun. Besides, there is an interesting phenomenon that the performance of P-CNN has a significant improvement compared to P\--CE detection. The reason may be attributed to that JPEG compression weakens the signal components in high frequence and the difference between original and enhanced images after JPEG compressing would be highlighted.

\begin{table}[!h]
	\footnotesize
	\caption{CE detection accuracy for pre-JPEG compressed images with different QFs. AVE is the average accuracy. Best results are marked in bold. }
	\label{tab1}
	\tabcolsep 10pt 
	\begin{tabular*}{\textwidth}{c|c|cccc|c}
		\toprule
		QF & Method & $\gamma=0.6$    &  $\gamma=0.8$  & $\gamma=1.2$ & $\gamma=1.4$ & AVE \\\hline
		
		\quad & De Rosa\cite{de2015second} &81.50\%  & 79.69\%  & 75.16\%  & 72.70\%  & 77.26\%\\
		\quad & Cao\cite{cao2014contrast} & 93.96\%  & 93.75\%  & 80.36\%  & 81.57\% & 87.41\%\\
		\quad & Li\cite{li2016identification} &  99.11\%  & 98.59\%  & 97.75\%  & 98.43\% & 98.47\%\\
		50 & Sun\cite{sun2018novel} & 99.73\%  & 99.62\%  & 99.40\%  & 99.75\% & 99.63\%\\
		\quad & P-CNN & 98.20\%  & 98.25\%  & 96.70\%  & 97.30\% & 97.61\%\\
		\quad & H-CNN &99.90\%  & 99.80\%  & 99.50\%  & 99.78\%  & 99.75\%\\
		\quad & DM-CNN & \textbf{99.97\%} & \textbf{99.90\%} & \textbf{99.86\%} & \textbf{99.96\%} & \textbf{99.92\%}\\ \hline
		
		\quad & De Rosa\cite{de2015second} & 83.99\%  & 82.27\%  & 77.47\%  & 72.95\% & 80.67\%\\
		\quad & Cao\cite{cao2014contrast} & 94.06\%  & 93.77\%  & 80.55\%  & 81.56\% & 87.49\%\\
		\quad & Li\cite{li2016identification} &  98.54\%  & 97.42\%  & 96.22\%  & 97.79\%  & 97.49\%\\
		70 & Sun\cite{sun2018novel} &99.32\%  & 99.12\%  & \textbf{99.14\%} & 98.89\% & 99.12\%\\
		\quad & P-CNN & 98.60\%  & 97.00\%  & 95.70\%  & 96.50\% & 96.95\%\\
		\quad & H-CNN & 98.86\%  & 99.03\%  & 98.27\%  & 97.68\% & 98.46\%\\
		\quad & DM-CNN & \textbf{99.68\%} & \textbf{99.51\%} & 99.06\% & \textbf{99.40\%} & \textbf{99.41\%}\\
		\bottomrule
	\end{tabular*}
\end{table}

For anti-forensic attacks, Cao's method does not work and there is a degradation in performance of H-CNN, especially, when anti-forensic method \cite{cao2010forensic} is applied. Because that the anti-forensic attacks would conceal the peak/gap feature in histogram domain. In addition, the anti-forensics attacks based on histogram maybe have a slight effect on pixel domain. Therefore, the P-CNN has better performance than H-CNN in this case. When the fusion framework is used to merge pixel and histogram domains together, DM-CNN obtained the best detection accuracy. While the pre-compression and anti-forensic attack are put into together, as shown in Table 4, the proposed CNN gains comparable performance with Li and Sun' scheme.

In conclusion, De Rosa's method is not robust for pre-JPEG compression and anti-forensics attack and Cao's method is vulnerable for anti-forenisic attack. Furthermore, such prior algorithms are unstable in different gamma levels. Although Li's method based on high dimensional features is better than previous works in the case of pre-JPEG compression and anti-forensic attack, its performance is unsatisfactory when no other operation is used. The deep learning-based method proposed by Sun obtained slight lower detection accuracy than the proposed DM-CNN, but it has a much higher computational cost during the feature extraction of the GLCM in preprocessing. Comparing with the above schemes, the proposed DM-CNN achieves good robustness against pre-JPEG compression, anti-forensic attacks, and CE level variation and obtains the best average detection accuracy in all cases studied.

\begin{table}[!h]
	\footnotesize
	\caption{CE detection accuracy in the case of anti-forensics attacks. $'-'$ denotes that the method does not work in this case. AVE is the average accuracy. Best results are marked in bold.}
	\label{tab1}
	\tabcolsep 10pt 
	\begin{tabular*}{\textwidth}{c|c|cccc|c}
		\toprule
		Attack & Method & $\gamma=0.6$    &  $\gamma=0.8$  & $\gamma=1.2$ & $\gamma=1.4$ & AVE \\\hline
		
		\quad & De Rosa\cite{de2015second} &61.67\%    &  58.83\%  & 55.32\% & 59.33\%  & 58.79\%\\
		\quad & Cao\cite{cao2014contrast} & $-$  & $-$    & $-$ & $-$ & $-$\\
		\quad & Li\cite{li2016identification} & 96.30\%  & 95.54\% & 95.72\% &96.55\% & 96.03\%\\
		\cite{cao2010forensic} & Sun\cite{sun2018novel} & 95.53\%  & 89.94\% & 90.55\% &92.42\% & 92.11\%\\
		\quad & P-CNN &\textbf{97.90\%}    &  \textbf{96.00\%} & 96.50\% & 96.55\%& 96.74\%\\
		\quad & H-CNN &88.77\%    &  73.65\%  & 74.85\% & 78.42\% & 78.92\%\\
		\quad & DM-CNN & 97.85\% & 95.97\% & \textbf{96.68\%} & \textbf{97.18\%} & \textbf{96.92\%}\\ \hline
		
		\quad & De Rosa\cite{de2015second} & 69.85\%    &  66.03\%  & 62.29\% & 64.42\%   & 65.65\%\\
		\quad & Cao\cite{cao2014contrast} & $-$  & $-$    & $-$ & $-$ & $-$\\
		\quad & Li\cite{li2016identification} &  99.57\%  & 99.38\% & 99.33\% &99.51\%  & 99.48\%\\
		\cite{barni2012universal} & Sun\cite{sun2018novel} &99.48\%    &  99.07\%  & 99.08\% & 99.19\% & 99.21\%\\
		\quad & P-CNN & 98.60\%    &  98.50\%  & 97.80\% & 98.00\%  & 98.21\%\\
		\quad & H-CNN & 98.82\%    &  97.59\%  & 97.57\% & 97.09\% & 97.77\%\\
		\quad & DM-CNN & \textbf{99.72\%} & \textbf{99.78\%} & \textbf{99.70\%} & \textbf{99.59\%} & \textbf{99.70\%}\\
		\bottomrule
	\end{tabular*}
\end{table}

\begin{table}[!h]
	\footnotesize
	\caption{CE detection accuracy for JPEG compressed images with different QFs and anti-forensics attack \cite{cao2010forensic}. $'-'$ denotes that the method does not work in this case. AVE is the average accuracy. Best results are marked in bold.}
	\label{tab1}
	\tabcolsep 10pt 
	\begin{tabular*}{\textwidth}{c|c|cccc|c}
		\toprule
		QF & Method & $\gamma=0.6$    &  $\gamma=0.8$  & $\gamma=1.2$ & $\gamma=1.4$ & AVE \\\hline
		
		\quad & De Rosa\cite{de2015second} & 70.26\%  & 67.85\%  & 65.38\%  & 66.52\%  & 67.50\%\\
		\quad & Cao\cite{cao2014contrast} & $-$  & $-$    & $-$ & $-$ & $-$\\
		\quad & Li\cite{li2016identification} &  99.90\%  & 99.90\%  & 99.90\%  & 99.90\%  & 99.90\%\\
		50 & Sun\cite{sun2018novel} &  99.75\%  & 99.63\%  & 99.68\%  & 99.57\% & 99.66\%\\
		\quad & P-CNN & 99.90\%  & 99.90\%  & 99.90\%  & 99.90\%  & 99.90\%\\
		\quad & H-CNN &99.45\%  & 99.40\%  & 99.20\%  & 99.20\%   & 99.31\%\\
		\quad & DM-CNN & \textbf{99.93\%} & \textbf{99.96\%} & \textbf{99.97\%} & \textbf{99.94\%} & \textbf{99.95\%}\\ \hline
		
		\quad & De Rosa\cite{de2015second} & 68.68\%  & 65.61\%  & 62.24\%  & 63.93\%  & 65.12\%\\
		\quad & Cao\cite{cao2014contrast} & $-$  & $-$    & $-$ & $-$ & $-$\\
		\quad & Li\cite{li2016identification} &  99.90\% & 99.90\%  & 99.90\%  & \textbf{99.90\%}   & 99.90\%\\
		70 & Sun\cite{sun2018novel} & 99.32\%  &99.34\%  & 98.60\%  & 99.03\%  & 99.07\%\\
		\quad & P-CNN & 99.80\%  & 99.75\%  & 99.55\%  & 99.80\%& 99.73\%\\
		\quad & H-CNN &  97.35\%  & 98.35\%  & 97.80\%  & 98.15\% & 97.91\%\\
		\quad & DM-CNN & \textbf{99.92\%} & \textbf{99.94\%} & \textbf{99.95\%} & \textbf{99.90\%} & \textbf{99.93\%}\\
		\bottomrule
	\end{tabular*}
\end{table}

\subsection{Exploration on the Strategy to Improve Performance of CNN-based CE Forensics}
Although numerous deep learning-based schemes have been proposed for digital image forensics, to the best of our knowledge, until now no one focus on exploring the strategy for performance improvement of single CNN-based CE forensics. However, it is important for the neophyte to design the new CNN architecture in the community of image forensics. In order to fill such gap, we make a preliminary exploration in this work. Specifically, there are three parts: adding the preprocessing, improving complexity of architecture, and selecting training strategy, which includes fine-tune technique and data augmentation.
\subsubsection{Preprocessing}
Through protracted and unremitting efforts of researchers, the deep learning technique developed for computer vision (CV) tasks has been succeeded in image forensics. Differing from CV related tasks, classification on image forensic has little relation to the image content. Therefore, preprocessing technique evolved into a universal way to improve the signal-to-noise ratio (SNR). High-pass filtering has become one of most popular means in preprocessing stage. In this part, using P-CNN in the case of $\gamma=0.6$ as an example, we evaluate six kinds of high-pass filters, H1, V1, H2, V2, LAP, HP, respectively, that widely applied into image forensics and compare them with the case without preprocessing. The definition of these filter are shown in Table. 5 and performance of the above cases is presented in Fig 7. $NON$ means the case without preprocessing. It can be seen that it is not good for CE forensic when non-preprocessing is used. In addition, first-order difference along horizontal direction has better performance. At the same time, the HP and LAP filter proposed for the other forensic task obtained worse performance, which indicates that it is necessary for image forensics to design different high-pass filters. 

\begin{table}[!h]
	\footnotesize
	\renewcommand\arraystretch{1.5}
	\caption{The filters evaluated in this work.}
	\label{tab1}
	\tabcolsep 15pt 
	\begin{tabular*}{\textwidth}{c|c|c}
		\toprule
		$ H1 = \begin{bmatrix}
		1 & -1
		\end{bmatrix}$ & $V1 = \begin{bmatrix}
		1\\
		-1 
		
	\end{bmatrix}$    & $ H2 = \begin{bmatrix}
		1 & 0\\ 
		0 & -1
	\end{bmatrix} $ \\\hline
	$	H2 = \begin{bmatrix}
		0 & -1\\ 
		1 & 0
	\end{bmatrix}$ & $LAP = \begin{bmatrix}
		0 & -1 & 0\\ 
		-1 & 4 & -1\\ 
		0 & -1 & 0
	\end{bmatrix}$ & $HP = \frac{1}{12}\cdot \begin{bmatrix}
		-1 & 2 & -2 & 2 & -1\\ 
		2 & -6 & 8 & -6 & 2\\ 
		-2 & 8 & -12 & 8 & -2\\ 
		2 & -6 & 8 & -6 & 2\\ 
		-1 & 2 & -2 & 2 & -1
	\end{bmatrix}$ \\
	
	\bottomrule
\end{tabular*}
\end{table}

\subsubsection{Powerful Convolutional Neural Networks}
Thanks to the development of deep learning technique in CV, more powerful CNNs (ResNet, XceptionNet, SENet) spring up at an increasing rate in recent years. However, because of the limitations in the forensics community, such as insufficient training dataset and hardware configuration, it would be difficult to evaluate all of them. In order to verify the effectiveness of powerful CNN in CE forensics, based on P-CNN, we replace its traditional convolutional layers with residual blocks that proposed in ResNet18. The result is shown in Fig 7. Comparing with the case of H1, detection accuracy of the Res\_H1 increases by 0.65\%. The above discussion, we make a conclusion that for CE forensics, powerful CNNs would enhance performance and preprocessing plays a more important role.   
\begin{figure}[h]
\centering
\includegraphics[width=10cm,height=6.5cm]{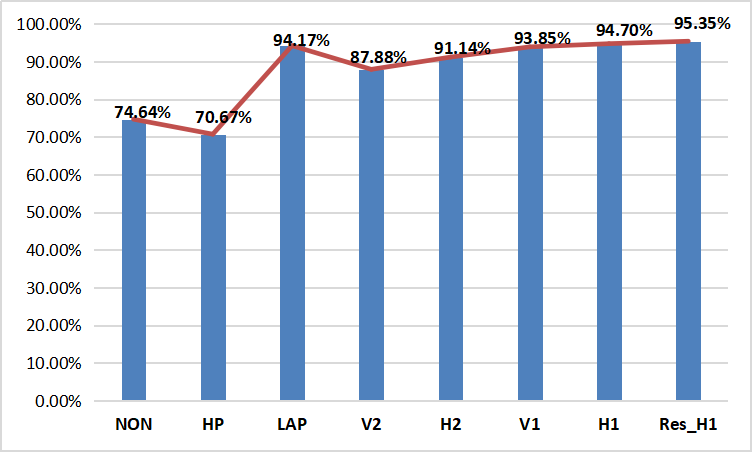}
\caption{Performance on P-CNN with/without preprocessing and with powerful network. NON means the case of P-CNN without preprocessing. The others represent the P-CNN with LAP, V2, H2, V1, H1 filter in the preprocessing, respectively. Res\_H1 denotes the P-CNN with H1 filter and residual blocks.}
\end{figure}
\subsubsection{Training Strategy}
It is well known that the scale of data has an important effect on performance for deep-learning based method and transfer learning technique \cite{pan2009survey} also provide an effective strategy to train the CNN model. In this part, we conducted experiments to evaluate the effect of the scale of data and  transfer learning  strategy, respectively, on performance of CNN. For the former, the images from BOSSBase are firstly cropped into 128x128 pixel patches with non-overlapping. Then these images are enhanced with $\gamma=0.6$. We randomly chose 80000 image pairs as test data and 5000, 20000, 40000, 80000 image pairs as training datas. Four groups of H-CNN, P-CNN are generated using above four training datas and the test data is same for these experiments. The result is as shown in Figure 8. It can be seen that the scale of training data has a slight effect on H-CNN with small parameters and the opposite happens for P-CNN. Therefore, larger scale of training data is beneficial to the performance of P-CNN with more parameters and the performance of P-CNN would be improved by enlarging training data. For the latter, we compare the performance of P-CNN with/without transfer learning in the cases of $\gamma=\{0.8, 1.2, 1.4\}$. The P-CNN with transfer learning by finetuning the model for $\gamma=\{0.8, 1.2, 1.4\}$ from the model for $\gamma=0.6$. As shown in Fig 9, P-CNN\--FT achieves better performance than P-CNN.

\begin{figure}[h]
\centering
\includegraphics[width=10cm,height=7cm]{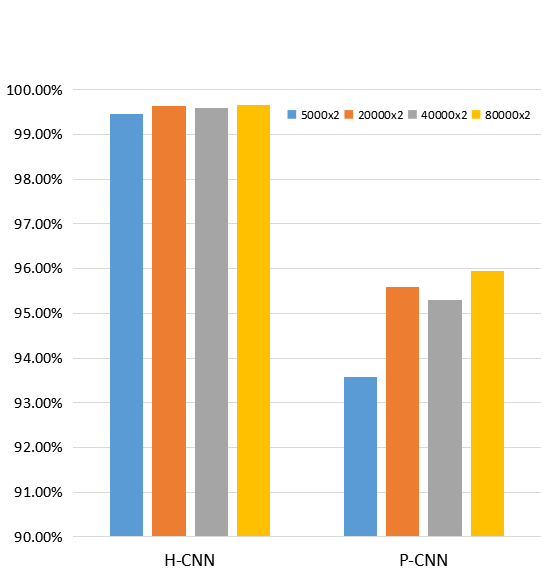}
\caption{Effect of the scale of training data.}
\end{figure}

\begin{figure}[h]
\centering
\includegraphics[width=10cm,height=6.3cm]{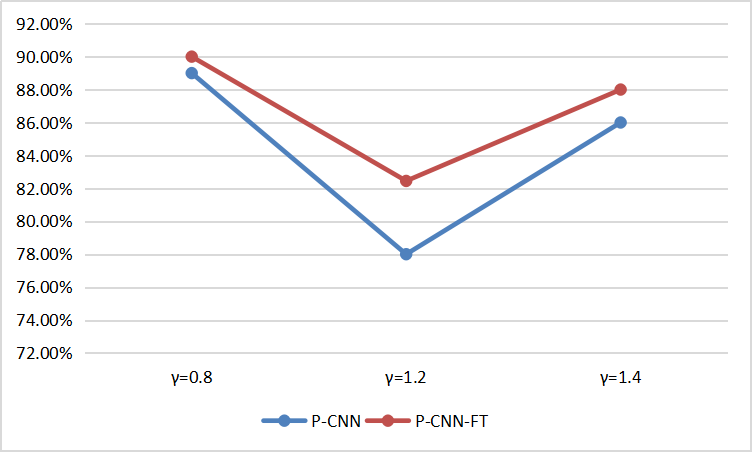}
\caption{Performance of the P-CNN and the P-CNN with fine-tune (P-CNN-FT).}
\end{figure}

\section{Conclusion}
The existing schemes for contrast enhancement forensics have an unsatisfactory performance, especially, in the cases of pre-JPEG compression and anti-forensic attacks. To deal with such problems, in this paper, a new deep learning-based framework dual-domain fusion convolutional neural networks (DM-CNN) is proposed. Such method achieve end-to-end classification based on pixel and histogram domains, which obtain great performance. Experimental results show that our proposed DM-CNN achieve better performance than the state-of-the-art ones and proposed method is robust against pre-JPEG compression, anti-forensic attack, and CE level variation. Beside, we explored on the strategy to improve performance of CNN-based CE forensics, which could provide guidance for the design of CNN-based forensics.

In sprite of good performance of exiting schemes, it is still a hard task to detect CE images in the case of post-JPEG compression with lower quality factors. The new algorithm should be designed to deal with this problem. In addition, the security of CNNs has drawn a lot of attention. Therefore, improving the security of CNNs is worth studying in the future.



\section{Acknowledgements}{This work was supported in part by the National Key Research and Development of China (No. 2016YFB0800404), the National NSF of China (Nos. 61672090, 61532005, 61332012, 61401408) and the Fundamental Research Funds for the Central Universities (Nos. 2018JBZ001, 2017YJS054). Pengpeng yang would like to acknowledge the CHINA SCHOLARSHIP COUNCIL, State Scholarship Fund, that supports his joint Ph.D program.
}

\section*{References}

\bibliography{mybibfile}

\begin{thebibliography}{10}
\expandafter\ifx\csname url\endcsname\relax
  \def\url#1{\texttt{#1}}\fi
\expandafter\ifx\csname urlprefix\endcsname\relax\def\urlprefix{URL }\fi
\expandafter\ifx\csname href\endcsname\relax
  \def\href#1#2{#2} \def\path#1{#1}\fi

\bibitem{stamm2008blind}
M.~Stamm, K.~R. Liu, Blind forensics of contrast enhancement in digital images,
  in: 2008 15th IEEE International Conference on Image Processing, IEEE, 2008,
  pp. 3112--3115.

\bibitem{stamm2010forensic}
M.~C. Stamm, K.~R. Liu, Forensic detection of image manipulation using
  statistical intrinsic fingerprints, IEEE Transactions on Information
  Forensics and Security 5~(3) (2010) 492--506.

\bibitem{stammConf}
M.~C. Stamm, K.~R. Liu, Forensic estimation and reconstruction of a contrast
  enhancement mapping, in: 2010 IEEE International Conference on Acoustics,
  Speech and Signal Processing, IEEE, 2010, pp. 1698--1701.

\bibitem{cao2014contrast}
G.~Cao, Y.~Zhao, R.~Ni, X.~Li, Contrast enhancement-based forensics in digital
  images, IEEE transactions on information forensics and security 9~(3) (2014)
  515--525.

\bibitem{li2016identification}
H.~Li, W.~Luo, X.~Qiu, J.~Huang, Identification of various image operations
  using residual-based features, IEEE Transactions on Circuits and Systems for
  Video Technology 28~(1) (2016) 31--45.

\bibitem{lin2013exposing}
X.~Lin, C.-T. Li, Y.~Hu, Exposing image forgery through the detection of
  contrast enhancement, in: 2013 IEEE international conference on image
  processing, IEEE, 2013, pp. 4467--4471.

\bibitem{lin2014two}
X.~Lin, X.~Wei, C.-T. Li, Two improved forensic methods of detecting contrast
  enhancement in digital images, in: Media Watermarking, Security, and
  Forensics 2014, Vol. 9028, International Society for Optics and Photonics,
  2014, p. 90280X.

\bibitem{wen2018contrast}
L.~Wen, H.~Qi, S.~Lyu, Contrast enhancement estimation for digital image
  forensics, ACM Transactions on Multimedia Computing, Communications, and
  Applications (TOMM) 14~(2) (2018) 49.

\bibitem{de2015second}
A.~De~Rosa, M.~Fontani, M.~Massai, A.~Piva, M.~Barni, Second-order statistics
  analysis to cope with contrast enhancement counter-forensics, IEEE Signal
  Processing Letters 22~(8) (2015) 1132--1136.

\bibitem{farid2001blind}
H.~Farid, Blind inverse gamma correction, IEEE Transactions on Image Processing
  10~(10) (2001) 1428--1433.

\bibitem{popescu2004statistical}
A.~C. Popescu, H.~Farid, Statistical tools for digital forensics, in:
  international workshop on information hiding, Springer, 2004, pp. 128--147.

\bibitem{cao2010forensic}
G.~Cao, Y.~Zhao, R.~Ni, Forensic estimation of gamma correction in digital
  images, in: 2010 IEEE International Conference on Image Processing, IEEE,
  2010, pp. 2097--2100.

\bibitem{wang2018parameter}
P.~Wang, F.~Liu, C.~Yang, X.~Luo, Parameter estimation of image gamma
  transformation based on zero-value histogram bin locations, Signal
  Processing: Image Communication 64 (2018) 33--45.

\bibitem{barni2012universal}
M.~Barni, M.~Fontani, B.~Tondi, A universal technique to hide traces of
  histogram-based image manipulations, in: Proceedings of the on Multimedia and
  security, ACM, 2012, pp. 97--104.

\bibitem{cao2010anti}
G.~Cao, Y.~Zhao, R.~Ni, H.~Tian, Anti-forensics of contrast enhancement in
  digital images, in: Proceedings of the 12th ACM Workshop on Multimedia and
  Security, ACM, 2010, pp. 25--34.

\bibitem{kwok2011alternative}
C.-W. Kwok, O.~C. Au, S.-H. Chui, Alternative anti-forensics method for
  contrast enhancement, in: International Workshop on Digital Watermarking,
  Springer, 2011, pp. 398--410.

\bibitem{comesana2013optimal}
P.~Comesana-Alfaro, F.~P{\'e}rez-Gonz{\'a}lez, Optimal counterforensics for
  histogram-based forensics, in: 2013 IEEE International Conference on
  Acoustics, Speech and Signal Processing, IEEE, 2013, pp. 3048--3052.

\bibitem{cao2014attacking}
G.~Cao, Y.~Zhao, R.~Ni, H.~Tian, L.~Yu, Attacking contrast enhancement
  forensics in digital images, Science China Information Sciences 57~(5) (2014)
  1--13.

\bibitem{ravi2015ace}
H.~Ravi, A.~V. Subramanyam, S.~Emmanuel, Ace--an effective anti-forensic
  contrast enhancement technique, IEEE Signal Processing Letters 23~(2) (2015)
  212--216.

\bibitem{barni2018cnn}
M.~Barni, A.~Costanzo, E.~Nowroozi, B.~Tondi, Cnn-based detection of generic
  contrast adjustment with jpeg post-processing, in: 2018 25th IEEE
  International Conference on Image Processing (ICIP), IEEE, 2018, pp.
  3803--3807.

\bibitem{zhang2018global}
C.~Zhang, D.~Du, L.~Ke, H.~Qi, S.~Lyu, Global contrast enhancement detection
  via deep multi-path network, in: 2018 24th International Conference on
  Pattern Recognition (ICPR), IEEE, 2018, pp. 2815--2820.

\bibitem{sun2018novel}
J.-Y. Sun, S.-W. Kim, S.-W. Lee, S.-J. Ko, A novel contrast enhancement
  forensics based on convolutional neural networks, Signal Processing: Image
  Communication 63 (2018) 149--160.

\bibitem{shan2019robust}
W.~Shan, Y.~Yi, R.~Huang, Y.~Xie, Robust contrast enhancement forensics based
  on convolutional neural networks, Signal Processing: Image Communication 71
  (2019) 138--146.

\bibitem{ioffe2015batch}
S.~Ioffe, C.~Szegedy, Batch normalization: Accelerating deep network training
  by reducing internal covariate shift, arXiv preprint arXiv:1502.03167.

\bibitem{yang2016recapture}
P.~Yang, R.~Ni, Y.~Zhao, Recapture image forensics based on laplacian
  convolutional neural networks, in: International Workshop on Digital
  Watermarking, Springer, 2016, pp. 119--128.

\bibitem{he2015spatial}
K.~He, X.~Zhang, S.~Ren, J.~Sun, Spatial pyramid pooling in deep convolutional
  networks for visual recognition, IEEE transactions on pattern analysis and
  machine intelligence 37~(9) (2015) 1904--1916.

\bibitem{mangai2010survey}
U.~G. Mangai, S.~Samanta, S.~Das, P.~R. Chowdhury, A survey of decision fusion
  and feature fusion strategies for pattern classification, IETE Technical
  review 27~(4) (2010) 293--307.

\bibitem{yang2017source}
P.~Yang, R.~Ni, Y.~Zhao, W.~Zhao, Source camera identification based on
  content-adaptive fusion residual networks, Pattern Recognition Letters.

\bibitem{fontani2013framework}
M.~Fontani, T.~Bianchi, A.~De~Rosa, A.~Piva, M.~Barni, A framework for decision
  fusion in image forensics based on dempster--shafer theory of evidence, IEEE
  Transactions on Information Forensics and Security 8~(4) (2013) 593--607.

\bibitem{Bossbasedata}
http://agents.fel.cvut.cz/stegodata/.

\bibitem{caffewebsit}
http://caffe.berkeleyvision.org.

\bibitem{pan2009survey}
S.~J. Pan, Q.~Yang, A survey on transfer learning, IEEE Transactions on
  knowledge and data engineering 22~(10) (2009) 1345--1359.

\end{thebibliography}

\end{document}